\documentclass[12pt]{article}
\usepackage{graphicx}
\usepackage{color}
\usepackage{amsmath,slashed, amssymb}
\usepackage{fancyhdr}
\usepackage{hyperref}
\usepackage[titletoc]{appendix}
\numberwithin{equation}{section} 

\def\beq{\begin{eqnarray}}
\def\eeq{\end{eqnarray}}
\def\bea{\begin{eqnarray*}}
\def\eea{\end{eqnarray*}}

\def\centeron#1#2{{\setbox0=\hbox{#1}\setbox1=\hbox{#2}\ifdim
\wd1\rangle\wd0\kern.5\wd1\kern-.5\wd0\fi
\copy0\kern-.5\wd0\kern-.5\wd1\copy1\ifdim\wd0\rangle\wd1
\kern.5\wd0\kern-.5\wd1\fi}}
\def\ltap{\;\centeron{\raise.35ex\hbox{$\langle$}}{\lower.65ex\hbox{$\sim$}}\;}
\def\gtap{\;\centeron{\raise.35ex\hbox{$\rangle$}}{\lower.65ex\hbox{$\sim$}}\;}

\newcommand{\newc}{\newcommand}
\newc{\qbar}{{\overline q}}
\newc{\Kahler}{Kahler }
\newc{\deltaGS}{\delta_{\rm GS}}
\begin{document}
\begin{titlepage}
\begin{flushright}
{\large SCIPP 16/09 \\
\large ACFI-T16-31 \\
}
\end{flushright}

\vskip 1.2cm

\begin{center}

\vskip 1.cm

%{\LARGE\bf Monodromy in QCD at large $N$:  $\theta$ and the $\eta^\prime$}
{\Large\bf $\theta$ and the $\eta^\prime$ in Large $N$ Supersymmetric QCD}

\vskip 1.cm

{Michael Dine$^{(a)}$, Patrick Draper$^{(b)}$,
Laurel Stephenson-Haskins$^{(a)}$, and Di Xu$^{(a)}$}
\\
\vskip 1.0cm
{\it $^{(a)}$Santa Cruz Institute for Particle Physics and
\\ Department of Physics, University of California at Santa Cruz \\
     Santa Cruz CA 95064  } \\
     ~\\
{\it $^{(b)}$Amherst Center for Fundamental Interactions, Department of Physics,\\ University of Massachusetts, Amherst, MA 01003
}\\
\vspace{0.3cm}

\end{center}

%\vskip 4pt

\vskip 1.0cm

\begin{abstract}

%Supersymmetry often provides control over strong dynamics.  We reconsider some longstanding questions in QCD, regarding $\theta$ periodicity and the $\eta^\prime$ in the large $N$ limit,
%as well as the role of instantons, examining these questions in the framework
%of models with small soft breaking of supersymmetry.  Instanton effects, when under control,
%are not exponentially suppressed at large $N$,  but known exact results are consistent with
%expectations from large $N$ perturbation theory.  Conjectured features of QCD at large $N$ are reproduced, including the existence of $N$ branches.			  Viewing real QCD as the limit of large soft breaking
%terms suggests the possibility of branched behavior, but also makes plausible behavior closer to that suggested by
%considering instanton effects with an infrared cutoff.  
%Deciding between branched or instanton possibilities
%in real QCD requires lattice experiments at large $N$.

We study the large $N$ $\theta$ dependence and the $\eta^\prime$ potential in supersymmetric QCD with small soft SUSY-breaking terms.  Known exact results in SUSY QCD are found to reflect a variety of expectations from large $N$ perturbation theory, including the presence of branches and the behavior of theories with matter (both with $N_f \ll N$ and $N_f \sim N$). However, there are also striking departures from ordinary QCD and the conventional large $N$ description: instanton effects, when under control,
are not exponentially suppressed at large $N$, and branched structure in supersymmetric QCD is always associated with approximate discrete symmetries. We suggest that these differences motivate further study of large $N$ QCD on the lattice.

\end{abstract}

\end{titlepage}

\section{Conjectured Behaviors of QCD at large $N$}

In \cite{witteninstantonsnot}, Witten suggested that instantons fail to provide even a qualitative
picture of the $\theta$ dependence of QCD and the solution of the $U(1)$ problem.
Instead, he advanced strong arguments that the large $N$ approximation was a much more useful tool.  Particularly remarkable
was his observation that in large $N$, the anomaly can be treated as a perturbation and the $\eta^\prime$
understood as a pseudogoldstone boson.

The large $N$ picture for the physics of $\theta$ and the $\eta^\prime$ rests on the assumption that correlation functions of $F \tilde F$ at zero momentum
behave with $N$ as
similar correlation functions {\it at non-zero} momentum in perturbation theory.
In particular, a Green's function with $n$ insertions of $F \tilde F$ behaves as 
\begin{align}
\langle \left(\int F\tilde F\right)^n\rangle\sim N^{-n+2}\;.
\label{eq:FFd}
\end{align}
With this assumption,
and the requirement of $2 \pi$ periodicity in $\theta$,
the vacuum energy must behave, to leading order in $1/N$, as 
\begin{align}
E(\theta)={\rm min}_k c \left ( \theta + 2 \pi k \right )^2 
\label{branched}
\end{align}
The minimization over $k$ reflects a branched structure in the theory, and ensures that $\theta$ is a periodic
variable~\cite{witteninstantonsnot,wittenetaprime2,witten1998}. The branches are characterized by a constant background topological charge density,
\begin{align}
\langle F\tilde F\rangle_k \propto (\theta+2\pi k)\;,
\label{eq:branches}
\end{align}
and are smoothly traversed under $\theta\rightarrow\theta+2\pi$.
A dual description of the branches in a higher dimensional gravity theory was analyzed in \cite{witten1998}.

With $N_f \ll N$, the fermions are expected to be a small perturbation of the large $N$ pure gauge theory.   In particular, the axial anomaly can be treated as a perturbation~\cite{wittenetaprime,wittenetaprime2}.  The mass of the $\eta^\prime$ is an ${\cal O}\left ({1 \over N}\right )$ effect, and a pseudo-Goldstone boson, the $\eta^\prime$, should be included in chiral perturbation theory in order to nonlinearly realize the approximate axial symmetry. To leading order in $1/N$ and in the chiral limit, its potential is obtained by the replacement 
\begin{align}
\theta \rightarrow \theta + {N_f \eta^\prime \over  f_\pi}
\label{eq:thetaeta}
\end{align}
 in the vacuum energy. This form is fixed by the axial anomaly.\footnote{Here $\eta^\prime/f_\pi$ is normalized as an ordinary angle, valued on $[0,2\pi)$.  In chiral perturbation theory, it is included at leading order in large $N$ by the substitution $\Sigma\rightarrow \Sigma e^{i\eta^\prime/f_\pi}$, where $\Sigma$ are the $SU(3)$ $\sigma$-model fields. The axial symmetry can be realized as $\eta^\prime\rightarrow \eta^\prime +\beta f_\pi$, and the anomaly coefficient is $N_f$, constraining the potential to have the form~(\ref{eq:thetaeta}). A different periodicity and anomaly are obtained if the $\eta^\prime$ is instead introduced with canonically normalized kinetic term.}  Including the branch label, 
\begin{align}
V_k(\eta^\prime)=c\cdot \Lambda^4 \left(\theta+2\pi k+\frac{N_f\eta^\prime}{f_\pi}\right)^2\;.
\end{align}
Taking $N_f=1$ as an example, under $\eta^\prime\rightarrow\eta^\prime+2\pi f_\pi$, the state passes
from one branch to another.
Because $f_\pi^2 \propto N$, the $\eta^\prime$ (mass)$^2$ is a $1/N$ effect.  Higher order interactions of the
$\eta^\prime$ are suppressed by powers of $N$, behaving as
\beq
V_n \sim \Lambda^4 N^2 \left ({\eta^\prime \over N f_{\pi}} \right )^n\;.
\label{wittenetaprimepotential}
\eeq
In other words, the $\eta^\prime$ a true Goldstone boson in the large $N$ limit, in the sense that its interactions
vanish rapidly as $N \rightarrow \infty$.

Note that $\theta$ can be absorbed into the $\eta^\prime$.  With at least one massless quark, and ignoring terms in the chiral lagrangian
associated with high scale (weak or above) physics, the $\eta^\prime$ potential has a minimum at the CP conserving
point.    
 
These expressions for $\theta$ dependence and the $\eta^\prime$ potential are in stark contrast
with qualitative expectations from instantons, assumed to be cut off in the infrared in
some manner.  In this case, one would expect a convergent Fourier series, for example, for $E(\theta)$
in the pure gauge theory:
\beq
E(\theta) = \Lambda^4 \sum_q c_q \cos(q\theta).
\label{instantontheta}
\eeq
Correlators of $n$ insertions of  $F \tilde F$ at zero momentum would scale with $N$
in a manner independent of $n$, i.e. the extra powers of $1/N$ expected from perturbation
theory counting at non-zero momentum would be absent. Likewise this picture makes a distinctive physical prediction for the couplings of the $\eta^\prime$:
the extra powers of $1/N$ in equation \ref{wittenetaprimepotential} should be absent.
We refer to behavior of the type of Eq.~(\ref{branched}) as ``monodromy" or ``branched"
behavior, while that of Eq.~(\ref{instantontheta}) as ``instanton" behavior. 

Lattice gauge theory is the only framework available in which the conjectured $\theta$ dependence of large $N$ QCD can be tested.
 However, such questions are technically extremely challenging. Some recent progress in testing Eq.~(\ref{eq:FFd}) was recently reported in~\cite{latticelargen}, but concrete tests of the predicted cuspy behavior near $\theta=\pi$, or the existence, lifetime, and other properties of the tower of $k$ branches, remains elusive.\footnote{See discussion in~\cite{teperreview}.} 

On the other hand, there are a variety of known theories that are similar but more tractable than QCD, including supersymmetric QCD (SQCD), deformed Yang-Mills, and QCD at large `t Hooft coupling, in which the $\theta$ dependence and existence of branches can be studied analytically~\cite{Shifman:1995ua,Evans:1997dz,konishiseibergwitten, witten1998, Unsal:2012zj,Poppitz:2012nz}. While differing in the details, these theories largely reflect the behaviors in Eqs.~(\ref{branched},\ref{eq:branches}). 

The case of SQCD will be analyzed in detail in this work. More generally, progress in the understanding of the dynamics of strongly coupled supersymmetric gauge theories~\cite{Affleck:1983mk,Seiberg:1994bz,intriligatorleighseiberg,Seiberg:1994pq} led to new studies of ordinary QCD, considering it as a limit of {\it Softly Broken Supersymmetric QCD} (SBQCD), or SUSY QCD with $N_f$ vectorlike flavors and soft SUSY-breaking masses~\cite{Aharony:1995zh, oz, Shifman:1995ua,Gabadadze:2002ff, martinwells}. We will study aspects of $\theta$ dependence in large $N$ SBQCD, including the existence of branches, $N$ scalings, the physics of the $\eta^\prime$, the role of instantons, and the sense in which adding matter can be thought of as a perturbation.  In Secs.~\ref{ndependence} and \ref{susybranches}, we observe a number of properties consistent with the large $N$ conjectures for ordinary QCD, including, as noted previously in~\cite{Shifman:1995ua,oz}, branched behavior (associated with the gaugino condensate in the SUSY limit, as well as $F\tilde F$ in the presence of soft breakings), and, as noted in~\cite{Aharony:1995zh,martinwells}, a supersymmetric version of the $\eta^\prime$ with mass of order $1/N$ in certain regions of parameter space.  

We also make several new observations. The behavior of ordinary QCD is different if the quark mass dominates
over the effects of the $U(1)_A$ anomaly and when the anomaly dominates.  In the former case, there are $N$
branches, while in the latter limit there are $N_f$ branches.  Phase transitions
are expected in passing between these regimes. In Sec.~\ref{phasestructure} we show that the same phenomenon arises in SBQCD
and we exhibit the phase structure.  In Sec.~\ref{matterpert} we demonstrate that small changes in the number of flavors $\Delta N_f\ll N$ leads to small changes in the physics of different vacua at large $N$: this provides a concrete realization of ``matter as a perturbation."  

Finally, we point out two ways in which the properties of SBQCD {\emph{differ}} from the conjectured properties of QCD. First, in Sec.~\ref{largeninstantons} we return to the fate of instantons in large $N$: the conjectured exponential suppression of instanton effects in QCD is critical to the large $N$ scaling properties described above. A simple heuristic argument
 suggests that if IR divergences associated with QCD instantons are cut off at a scale of order $\Lambda_{QCD}^{-1}$, there
 is no exponential suppression.  As a counterargument, Ref.~\cite{witteninstantonsnot} emphasized
 that because of the extreme nature of the power
 law divergences, the result is extremely sensitive to how the cutoff is chosen, and the notion that such
 a cutoff computation makes sense, even at a qualitative level, is hard to support.  But in SQCD with $N_f = N-1$, where a systematic instanton computation of holomorphic quantities is possible, we show that the results are not suppressed by $e^{-N}$, and that the gauge boson mass acts as an infrared cutoff approaching $\Lambda$ at precisely the required rate.
 On the other hand, the $N$-scalings are, in fact, exactly as predicted by perturbative arguments,
 and the $\theta$-dependence reflects the branched structure!
 We provide other evidence, in less controlled situations, that a notion of cut-off instantons may survive in supersymmetric theories in large $N$.

Secondly,  in Sec.~\ref{realqcd}, we comment on the role of discrete symmetries. Unlike QCD, branched structure in SBQCD is associated with an approximate $Z_N$ symmetry, and a corresponding set of $N$ quasi-degenerate, metastable ground states. What happens to these states in the limit of large soft breakings, where the discrete
symmetry is lost and QCD is recovered?  A priori, one possibility is that these states, and
the associated branch structure, disappears. The possibility of phase transitions
as parameters are varied is already realized in supersymmetric QCD in the controlled approximation
of small soft breakings.
 Against this possibility is the
usual large $N$ scaling of perturbative correlation functions,  suggesting that the branches should remain.  As we briefly review, a possible microscopic realization of
the branches in real QCD is provided by 't Hooft's proposal of {\it oblique confinement}~\cite{oblique} (particularly as realized in deformed $N=2$ theories).\footnote{We thank Ed Witten and
Davide Gaiotto for stressing this possibility to us.} On the other hand,
the fact that instantons are not suppressed as $e^{-N}$ in controlled situations raises questions
about these arguments.    Whether the states disappear or survive cannot be conclusively established without non-perturbative computations.

In Sec.~\ref{conclusions}, we summarize and conclude. We argue that while the traditional large $N$ branched picture of~\cite{witteninstantonsnot,wittenetaprime,wittenetaprime2,witten1998} remains likely, only lattice calculations can ultimately settle the issues. 
%We conclude with a brief discussion of directions for future work on this question.

 \section{Large N Scaling of the Gaugino Condensate}
 \label{ndependence}
 
 Much is understood about the dynamics of supersymmetric gauge theories.  For a pure supersymmetric
 gauge theory, for example, the value of the gaugino condensate is known, from arguments which
 resemble neither perturbation theory nor a straightforward instanton computation~\cite{Novikov:1983ee,Novikov:1983uc,Affleck:1983mk,Novikov:1985ic,intriligatorleighseiberg,Finnell:1995dr}.
 It is interesting that, as we now show, the $N$ dependence agrees with that expected from the usual diagrammatic counting.

Let us recall the Coleman-Witten argument~\cite{colemanwitten} for the $N$-scaling of the chiral condensate in QCD and  apply it to supersymmetric QCD. 
By ordinary $N$ counting, an effective potential for ${\cal M}=\langle \bar \psi \psi \rangle$  (with $\psi, ~\bar \psi$ two-component fermions) would take the form
 \beq
 V({\cal M}) = N f\left({{\cal M}^\dagger {\cal M} \over N^2\Lambda_{QCD}^6}\right)\;,
 \eeq
 in the fermion normalization where $1/g^2$ sits in front of the whole action.
Thus ${\cal M} \propto N \Lambda^3.$  
 For supersymmetric gauge theories, the corresponding analysis for
 the $\langle \lambda \lambda \rangle$ effective potential gives
 \beq
 V(\langle \lambda \lambda \rangle) = N^2 f\left({\langle \lambda \lambda \rangle\langle \lambda \lambda \rangle^*
 \over N^2\Lambda^6}\right)
 \eeq 
 again in the gaugino normalization where $1/g^2$ sits in front of the whole action. 
 So, we expect $\langle \lambda \lambda \rangle = N\Lambda^3$.

The exact result in pure gauge theory is
 \begin{align}
\langle \lambda \lambda \rangle = 32 \pi^2 \Lambda_{hol}^3e^{\frac{2\pi i k}{N}}\;.
\label{eq:llvev}
\end{align}
(For a review, see~\cite{peskinduality}.) Here $\Lambda_{hol}$ is the holomorphic $\Lambda$ parameter, proportional to $e^{\frac{i\theta}{3N}}$.  In general, as discussed in \cite{shifmannscaling},
the holomorphic $\Lambda$ parameter differs from the more conventional $\Lambda$ parameter, as defined in \cite{pdgqcd},
 by an $N$-dependent factor:
 \beq
 \Lambda_{hol} = \Lambda \left ({b_0 \over 16 \pi^2} \right )^{b_1/b_0^2}\;.
 \label{eq:lambdarel}
 \eeq
 We review this connection in Appendix A. Eq.~(\ref{eq:lambdarel}) reflects the fact that $\Lambda$ is fixed as $N \rightarrow \infty$ with $g^2 N$ fixed, while $\Lambda_{hol}^3\propto N \Lambda^3$.
It is striking that the  $N$ scaling of $\langle \lambda \lambda \rangle$ agrees with the diagrammatic expectation, although the physics leading to the exact computation appears quite different.

 \section{$\theta$ and the $\eta^\prime$ Potential in SQCD}
 \label{susybranches} 
 
 In this section, we will see that with small soft breakings, both without matter and with $N_f \ll N$,
 supersymmetric theories exhibit precisely the branched behavior anticipated by Witten,
 with the branches being associated with the breaking of an approximate discrete symmetry.
 
 \subsection{Supersymmetric $SU(N)$ Gauge Theory Without Matter}

 For vanishing gaugino mass, the gaugino condensate is given by Eq.~(\ref{eq:llvev}).
% \beq
% \langle \lambda \lambda \rangle = 32\pi^2 \Lambda_{hol}^3 e^{2 \pi i k \over N},
% \eeq
% where, in the case of the theory without matter, 
% $
% \Lambda_{hol}^3 \propto N \Lambda_{QCD}^3.
% $
 In the presence of a small holomorphic soft-breaking mass, $m_\lambda$, the vacuum energy is
 \beq
 V(\theta,k) \simeq  {m_\lambda} \vert\Lambda_{hol}\vert^3  \cos \left({\theta + 2 \pi k \over N}\right).
 \eeq
 In terms of physical quantities, 
 \beq
 m_\lambda \Lambda_{hol}^3 = N^2 m_{phys} \Lambda^3,
 \eeq
 where  $m_{phys} = g^2 m_\lambda$. Therefore,
for very large $N$ with $\theta$ and $k$ fixed
%, noting that the physical gaugino mass satisfies
%, we can expand:
 \beq
 V(\theta,k)\simeq N^2 m_{phys} \vert\Lambda\vert^3  \left({\theta + 2 \pi k \over N}\right)^2.
 \eeq
 This is compatible with the $N$-scaling and $\theta$ dependence of~\cite{witten1998}. 
 
 For small $m_\lambda$, the separate branches are long-lived.  As $m_\lambda$ increases, approaching real QCD, the fate of the branches is not clear; we will comment on this further in Sec.~\ref{realqcd}.

 \subsection{$N_f \ll N$ in supersymmetric QCD:  A model for the $\eta^\prime$}
 \label{etaprime}
 
 Supersymmetric QCD with $N_f<N$ flavors possesses an $SU(N_f)_L \times SU(N_f)_R \times U(1)_B\times  U(1)_R$ symmetry.
Dynamically, a non-perturbative superpotential is generated~\cite{Affleck:1983mk},
 \beq
 W_{np} = (N-N_f){ \Lambda_{hol}^{3N - N_f \over N-N_f} \over (\det \bar Q Q)^{1 \over N-N_f}}.
 \eeq
Including supersymmetric mass terms for the quarks, the system has $N$ supersymmetric vacua.
 
 Turning on general soft breakings gives a set of theories which, in certain limits, should reduce to $SU(N)$ QCD with $N_f$ flavors of
 fermionic quarks.  For small values of the supersymmetric mass terms and the soft breaking terms,
 the system can be studied in a systematic perturbative/semiclassical approximation~\cite{Aharony:1995zh,martinwells}.  Consider first adding
only soft squark and gaugino masses:
 \beq
 \delta V = \tilde m^2 \sum_f \left ( \vert Q_f \vert^2 + \vert \bar Q_f \vert^2 \right )+ m_\lambda \lambda \lambda.
 \label{softsquarkmasses}
 \eeq
With universal soft scalar mass terms, the first terms respect the full  $SU(N_f)_L \times SU(N_f)_R \times U(1)_B\times  U(1)_R$
symmetry of the supersymmetric theory.  The gaugino mass term breaks the $U(1)_R$.

Ignoring  the gaugino mass, the potential
\beq
V = \sum_f \left (\vert {\partial W \over \partial Q_f }\vert^2 + \vert {\partial W \over \partial \bar Q_f }\vert^2 \right )
+ \delta V
\eeq
(along with the $\sum (D^{a})2$ terms) yields a minimum at
\beq
Q^a_f = v \delta^a_f~~~\bar Q^a_f = Q^a_{f^\prime} U_{f^\prime f}\;,
\label{qvevs}
\eeq
where $U$ is a unitary matrix  describing the Goldstone fields.  
If $U=1$, the symmetry is broken to the diagonal subgroup.  
$v$ is given, in the large $N$ limit, by:
\beq
v =\Lambda_{hol}  \left ({\Lambda_{hol}^2 \over \tilde m^2} \right )^{1/4}.
\label{vvalue}
\eeq
(If we take $\tilde m^2 \sim \Lambda^2$, and recall that $\Lambda_{hol}^3 \sim N \Lambda^3$,
then $v = f_{\eta^\prime} \sim \sqrt{N}$, as expected by standard large $N$ arguments. The same result is obtained  if the moduli are stabilized by a small quark mass, $v^2\sim \Lambda_{hol}^3/m\Rightarrow v\sim \sqrt{N}$.)

The gaugino bilinear $\lambda \lambda$ has an expectation value in this theory,
which is essentially the derivative with respect to $\tau$ of the expectation value of the non-perturbative
superpotential~\cite{intriligatorleighseiberg,peskinduality},
\beq
\langle \lambda \lambda \rangle = {32 \pi^2}  \left \langle{ \Lambda_{hol}^{3N - N_f \over N-N_f} \over  (\det \bar Q Q)^{1 \over N-N_f} } \right \rangle.
\eeq
To leading order, the expectation value is obtained simply using the value of $v$ in Eq.~(\ref{vvalue}).
For large $N$, the condensate behaves as
\beq
\langle \lambda \lambda \rangle = \Lambda_{hol}^3 e^{{2 \pi i k \over N}+ i\,{\rm arg} \det U^{1/N}},
\eeq
where $U$ is the unitary matrix in Eq.~(\ref{qvevs}).

Now consider turning on a small $m_\lambda$.  The gaugino mass breaks the classical, anomalous $U(1)_R$ as well as the quantum, non-anomalous $U(1)_R$.
It also breaks the quantum $Z_N$ symmetry.  Through a 
field redefinition, we can take $m_\lambda = \vert m_\lambda \vert e^{i\theta/N}$.  Gaugino condensation then generates a potential for the fields $U$, which at large $N$ takes the form:
\beq
V(\theta,\eta^\prime) =  \vert m_\lambda \vert \Lambda_{hol}^3 \cos\left({\theta + 2 \pi k + {\eta^\prime \over v} \over N} \right),
\label{smallmlambda}
\eeq
where we have written $\arg \det U = {\eta^\prime \over v}$.
Recall that in conventional large $N$ scaling, $m_\lambda \propto N m_{\lambda}^{phys}$, where
$m_{\lambda}^{phys}$ is the physical gaugino mass.
Therefore, expanding for very large $N$ and taking $m_{\lambda}^{phys} \sim \Lambda$ gives the potential for the $\eta^\prime$ proposed in~\cite{wittenetaprime2}.  The scaling with $N$ is exactly as predicted.

For zero supersymmetric quark mass,  $\theta$ and $k$ can be removed by a redefinition of the $\eta^\prime$ field.  In the presence of a quark mass term, this is no longer the case.
The $\eta^\prime$ potential contains an additional term, which at large $N$ takes the form
\beq
V(\theta,\eta^\prime) =  \vert m_\lambda \vert \Lambda_{hol}^3 \cos\left({\theta + 2 \pi k + {\eta^\prime \over v} \over N} \right) + \vert m_q \vert\Lambda_{hol}^3
\cos
\left(\frac{\eta^\prime}{v} + \beta\right),
\label{smallmlambdamq}
\eeq
where $\beta$ is the phase of the quark mass. We comment on the properties of this potential in Sec.~\ref{phasestructure}.

\section{Phases with General $N_f<N$}
\label{phasestructure}

In QCD, the realization of branched structure is thought to vary with $m_q$~\cite{wittenetaprime2}.  At zero $m_q$, a field redefinition can eliminate $\theta$-dependence.  At large $N$, this corresponds to the fact that $\theta$ can be eliminated
by a shift of the $\eta^\prime$.  On the other hand, at sufficiently large $m_q$, the quarks can be integrated out and $\theta$-dependence
should reappear, along with any branched structure.

In SBQCD, 
already in the limit of soft breakings,  an intricate phase structure arises by varying the
soft breaking parameters and the quark masses.  This can be anticipated because in the theory of Eq.~(\ref{softsquarkmasses}), before including the quark masses $m_q$, the discrete symmetry is $Z_{N_f}$, a preserved subgroup of the anomalous $U(1)_A$ axial symmetry acting on $Q,\bar Q$.
If we set $m_\lambda$ to zero, with non-zero $m_q$, the discrete symmetry is $Z_N$, a preserved subgroup of the anomalous $U(1)_R$ symmetry acting only on $\lambda$.  It is easy to check that
varying the parameter
\beq
x = {m_\lambda \over m_q}\;,
\eeq
the number of local minima of the potential changes from $N$ at small $x$ to $N_f$ at large $x$.  

To see this explicitly, take the simplified case $\vert m_q \vert^2, \vert m_\lambda \vert^2 \ll \tilde m^2$, and $\tilde m^2$, $m_q$ proportional
to the unit matrix in flavor space.  We can then take $\bar Q Q = v_0^2 e^{i \eta^\prime}$ (note here we are working with a dimensionless $\eta^\prime$).  The potential
for the $\eta^\prime$ then has the form:
\beq
V(\eta^\prime) = m_q \Lambda_{hol}^{3N-N_f \over N-N_f} v_0^{-{2N_f \over N-N_f}} \cos\left({N \over N-N_f} \eta^\prime\right)
+  N m_\lambda {\Lambda_{hol}^{3N-N_f \over N-N_f} \over v_0^{2N_f \over N-N_f} } \cos\left( \eta^\prime {N_f \over N-N_f}\right),
\eeq
or, for $N \gg N_F$, 
\beq
V(\eta^\prime) = m_q \Lambda_{hol}^{3} v_0^{-{2N_f \over N}} \cos( \eta^\prime)
+  N m_\lambda \Lambda_{hol}^{3}  \cos\left( \eta^\prime {N_f \over N}\right).
\eeq
This potential is similar in structure to that for the ordinary $\eta^\prime$ proposed in~\cite{wittenetaprime2}.
It exhibits $N$ vacua in the limit of small $x$, and $N_f$ in the limit of large $x$. Analogously, in ordinary QCD, the large-$N$ $\eta^\prime$ potential has $N_f$ vacua in the limit $m_q\ll \Lambda/N$, and $N$ vacua in the opposite limit.

In SQCD, the transitions between these phases occur for $x$ of order one.  As the vacua disappear, they become increasingly unstable.
In the limit of large $x$, correlation functions with successively more insertions of $\int d^4 x F \tilde F$ are suppressed
by $N_f$, not $N$.  The potential can also be analyzed in the case of $N_f = N-1$, where a reliable instanton computation is possible. In this case, there are of order $N$ branches in either limit, but one can still observe transitions between different phases, increasing confidence in the small $N_f$ analysis.

The phase structure also offers some insight into the lifetimes of states of a system as one approaches the critical values
$x_0$ where they disappear.  The bounce action vanishes as a power of $x-x_0$ (of course, the semiclassical analysis
breaks down once the lifetime becomes short).

\section{Matter as a Perturbation}
\label{matterpert}
In the large $N$ limit, we might expect that small changes in the number of flavors  only affect the properties of the theory at order $1/N$: in this sense, matter is a perturbation. 

There are two classes of quantities we might study.  In actual QCD, we might ask about the $N_f$ dependence of the glueball mass or $F \tilde F$ correlation functions, expecting weak sensitivity of these quantities to ${\cal O}(1)$ changes in $N_f$ at large $N$.  Alternatively, we can consider the structure
of the quark sector.  Here we expect the features of the effective
action for the $\eta^\prime$, for example, to be determined by the large $N$ pure gauge theory.

In the supersymmetric theories, the gluino condensate is in the first class, and we expect small changes in the number of flavors
to yield only small changes in the condensate.  To test this idea, we must be precise about what is perturbed.  As we vary $N_f$, we hold the ultraviolet cutoff $M$ and the gauge coupling $g^2(M)$ fixed.  For simplicity, we take all quarks
to have mass $m_q$, with $m_q \gg \Lambda$, and we study the Wilsonian effective action at a scale $\mu$ such that $m_q \gg\mu \gg \Lambda$. Integrating out the quarks generates a term
\beq
{\cal L} = -{1 \over 32 \pi^2} \int d^2 \theta \left ({8 \pi^2 \over  g^2} + 3N \log(\mu/M) - N_f \log(m_q/M) \right )W_{\alpha}^2 .
\eeq
From it, we can compute the holomorphic low energy scale, 
$\Lambda_{LE}$, which in turn determines $\langle \lambda \lambda \rangle$,
\beq
\langle \lambda \lambda \rangle = \Lambda_{LE}^3 = \Lambda^3\left( {m_q \over \Lambda}\right)^{N_f \over N}.
\eeq
This expression is clearly smooth with respect to changes in $N_f$. Indeed, we can treat an additional flavor as a perturbation, computing
first the change in the effective action, and from that the change in $\Lambda_{LE}$.

Alternatively, we can consider a quantity involving the quark superfields, for small number 
of flavors.  
As before, we can think
of a fixed cutoff scale and coupling, and take universal quark masses $m_q \gg \Lambda$.  Then
integrating out the heavy fermions yields
\beq
\langle \bar Q Q \rangle = {1 \over 16 \pi^2 m_q} \langle \lambda \lambda \rangle\;.
\eeq
(this is an example of the Konishi anomaly \cite{konishianomaly}).
This agrees with the exact result, and by holomorphy, it holds for all $m_q$. Thus for small $N_f$ the quark condensate is determined in large $N$ by the pure gauge theory.  One can provide a heuristic derivation of this result at small $m_q$ as well.

For larger values of $N_f$, small changes $\Delta N_f\ll N$ should also produce only small changes in the theory, for appropriate choices of ground states. This is particularly interesting for $N_f=N-2, N-1, N, N+1, N+2$, where  the dynamics, when the quarks are light, is substantially different in each case (described via gaugino condensation, instantons, the deformed moduli space, s-confinement, and Seiberg duality, respectively~\cite{Affleck:1983mk,Seiberg:1994bz,Seiberg:1994pq}.) Yet, in large $N$, all descriptions must in some sense converge, up to $1/N$ corrections!

Let us understand a few simple reflections of this fact, again taking $m_{f\bar f}\rightarrow m\delta_{f\bar f}$ and $Q\bar Q_{f\bar f}\rightarrow v^2\delta_{f\bar f}$. For $N_f=N-1$, there is a Wilsonian effective superpotential~\cite{Affleck:1983mk},
\begin{align}
W_{Wilsonian}=\frac{\Lambda^{2N+1}}{v^{N-1}}+N_f m v^2
\end{align} 
and the vacuum is 
\begin{align}
v=\Lambda\left(\frac{\Lambda}{m}\right)^\frac{1}{2N}
\end{align}
which approaches $v\rightarrow\Lambda$ in the large $N$ limit, losing its $m$-dependence. In contrast, the case $N_f=N-1$ has a deformed moduli space~\cite{Seiberg:1994bz}, described by a 1PI effective superpotential with Lagrange multiplier $X$,
\begin{align}
W_{1PI}=X\left(v^{2N}-B\bar B-\Lambda^{2N}\right)+N_f mv^2\;.
\end{align}
Since there are no baryonic operators in $N_f=N-1$, vacua on baryonic branches are not connected to vacua in $N_f=N-1$. The meson vacuum, however, is: in the large $N$ limit, the $N_f=N-1$ vacuum becomes the $B=\bar B=0$ vacuum $v=\Lambda$ of $N_f=N$. The gaugino condensates likewise match in large $N$, and vanish in the massless limit.

A similar result is obtained for $N_f=N+1$ with small quark mass: the meson vev takes the form $v^{2N+1}=m\Lambda^{2N}$, so $v\rightarrow \Lambda$ in large $N$. The new feature of the $N_f=N+1$ theory, the chiral preserving vacuum, is obtained in the limit $m\rightarrow 0$, which does not commute with $N\rightarrow\infty$.

\section{Instantons at Large $N$}
\label{largeninstantons}

We see that approximately supersymmetric theories exhibit many of the features anticipated for real QCD, within
controlled approximations.  Much of our understanding of supersymmetric dynamics, on the other hand, involves
instantons in an essential way. 
This suggests that instanton effects {\it are not necessarily} suppressed at large $N$, and can
have controlled large $N$ limits, at least in SQCD.

\subsection{Heuristic treatment of instantons:  the infrared cutoff}
\label{doinstantonslie}

In the introduction, we discussed two potential behaviors for large $N$ QCD as a function of $\theta$, referred to as branched and instanton behaviors, respectively.  We have seen that supersymmetric SQCD with small gaugino mass exhibits the former
behavior.
Ref.~\cite{witteninstantonsnot} offered a simple argument against the latter, suggesting
that instanton effects are exponentially
suppressed in large $N$.  Let us recapitulate the argument. 

Consider QCD without flavors.  The one-instanton contribution to $V(\theta)$ has the structure:
\beq
V(\theta) = \int d\rho \rho^{-5 + {11 N \over 3}} M^{11 N \over 3} N e^{-{8 \pi^2 \over g(M)^2}} \cos(\theta)
\eeq
where $M$ is a renormalization scale.  Since $g^2(M) \sim 1/N$, this is formally exponentially
suppressed, but the expression is also infrared divergent.
Suppose that the integral is cut off at $\rho \approx \Lambda^{-1}$.  The result would then be simply
\beq
V(\theta) = {\rm C} \Lambda^4 \cos(\theta).
\eeq
which is of order one in large $N$.  Of course, this argument is handwaving at best.  If the cutoff
is $c~ \Lambda$, with $c$ an order one constant, then the result can be exponentially suppressed or enhanced by
$c^N$. Ref.~\cite{witteninstantonsnot} suggested that the most likely smooth limit for instanton effects in large $N$ is zero.

Imagine, however, that $c$ approaches $1$ as $e^{1/N}$: in this case, the limit of the single instanton term would be smooth and finite. In QCD, such a picture could only be qualitative; perturbative corrections and instanton-antiinstanton corrections are
all be nominally of the same order, and a reliable semiclassical calculation is not possible.   The only statement one could make, in general, is that $\theta$ dependence
would be described by
a series of the form of Eq.~(\ref{instantontheta}).
One could speculate on the convergence of the series, for example whether cusps arise in the potential.  This appears to occur in the $CP^N$ models, where finite temperature provides an
infrared cut-off on instanton size
\cite{affleck1,affleck2,affleck3}, and the series (\ref{instantontheta}) exhibits cusps in the limit $T \rightarrow 0$ (the Fourier expansion for $dE \over d\theta$
does not converge).  This will be discussed more fully in a subsequent publication.

\subsection{Scaling of Reliable Instanton Computations with $N$}
In SQCD with $N_f=N-1$, the role of instantons in large $N$ can be assessed sharply, exploiting the existence
of a pseudomoduli space. The effective superpotential can be computed systematically, and infrared divergences are cut
off by $Q \bar Q_{f\bar f} \equiv v^2\delta_{f\bar f}$.  The $\rho$ integrals take the form
\begin{align}
W \;\sim\; \int d\rho \,{(\Lambda \rho)}^{2N+1} {(v^*)}^{2N-2} \rho^{4N-5} e^{-c ^2\rho^2 \vert v \vert^2}\;\sim\; %c^{-4N+4}
 {\Lambda^{2N+1} \over v^{2N-2}}.
 \end{align}
A careful analysis yields~\cite{intriligatorleighseiberg}
\beq
W = {\Lambda_{hol}^{2N + 1} \over \det{\bar Q Q} }\;,
\eeq
which is na\"ively of order $e^{-N}$.

However, $v^2$ also depends on $\Lambda$.  For simplicity, taking all of the quarks to have equal mass,
\beq
v^N = \Lambda_{hol}^{N} \left ( {\Lambda_{hol} \over m_q} \right)^{1 \over N}.
\label{vequation}
\eeq
At the stationary point,
\beq
\langle W \rangle = a \Lambda_{hol}^2 m_q \left [ {\Lambda_{hol} \over m_q} \right]^{1/N}. 
\label{Wvev}
\eeq
This structure is dictated by symmetries and holomorphy.  In particular, there is a non-anomalous, spurious $R$ symmetry
under which
\beq
m_q \rightarrow e^{2 i \alpha{N \over N_f} }m_q.
\eeq
Similarly, there is a non-anomalous $R$ symmetry under which $m_q$ (and $Q,\bar Q$) are neutral, and $\Lambda \rightarrow e^{i \alpha 
2N/(2N+1) } \Lambda$. 

Eq.~(\ref{Wvev}) is notable.  First, there is no exponential suppression with $N$:  
\beq
\Lambda_{hol}^{2 + {1 \over N}} = M^{2+ {1 \over N}} e^{-{8 \pi^2 \over g^2 N}+ i {\theta \over N}}.
\eeq
Not only do the $e^{-{8 \pi^2 \over g^2}}$ factors appear with a suitable power to avoid $e^{-N}$ suppressions, but there are no
factors like $\pi^N$ or $2^N$ which might have obstructed a suitable large $N$ limit.  At the same time, the result exhibits monodromy,
arising from the $N$ roots of Eq.~(\ref{vequation}).

It is also important to stress that, unless $m_q$ is exponentially small, the stationary point lies in a region of strong coupling.
So a reliable calculation is possible taking $m_q = \epsilon^N \Lambda$, for small $\epsilon$, and then
using holomorphy and symmetries to extend the result to $m_q=\Lambda$.  For $m_q \sim \Lambda$, the 
instanton result is not reliable in the sense that non-holomorphic quantities like the scalar potential are not properly computed.  But the result for $\langle W\rangle$ qualitatively
has the instanton structure, and it is equivalent to say that it is saturated by the single instanton.  

We also note that in presence of a gaugino mass, we again find the usual formula for the vacuum energy,
\beq
E(\theta) = m_\lambda\langle W \rangle = m_\lambda \Lambda_{hol}^3 \cos\left({\theta+2 \pi k \over N}\right).
\eeq
So in this case, we have \emph{complete agreement} with expectations based on $N$ counting of perturbative Feynman diagrams, yet the result arises entirely from an instanton!
In particular, correlators of $n$ $F \tilde F$ operators at zero momentum behave as $N^{2-n}$, precisely as expected.    We have already noted how a cutoff might approach $\Lambda$ in large $N$ so that instanton amplitudes are unsuppressed. Here we see that, in the nearly supersymmetric case, the $\Lambda$ which
appears in the argument is the {\it holomorphic} $\Lambda$, yielding $\cos(\theta/N)$.  

To summarize, on the one hand, we see evidence for a branched structure, a structure originally suggested by a presumed suppression
of instanton effects.  On the other hand, we see that instantons are not suppressed, and the branches
are associated with an approximate discrete symmetry.  We cannot draw conclusions about the fate of the branched structure as 
SUSY breaking is increased, but the instanton argument for the branched structure, by itself, is at least misleading in the nearly-SUSY limit.  

\subsection{Further circumstantial evidence for the role of instantons}

Also instructive are instanton computations in the pure supersymmetric gauge theory.   This subject was pioneered
in \cite{shifmanvainshtein1,shifmanvainshtein2}.  In pure $SU(N)$ supersymmetric QCD, one can attempt to calculate the correlation function
\beq
G^{2N}=\langle \lambda \lambda(x_1) \dots \lambda \lambda(x_N) \rangle\;.
\eeq
A single instanton makes an infrared finite contribution to this correlator,
$G^{2N} \sim \Lambda^{3N}$, which is formally of order $e^{-N}$.    This paper argued that this correlation function, as the correlator
of the lowest component of a set of chiral fields, was independent
of coordinates, and in addition advanced arguments that it was not renormalized. The authors of \cite{rossiveneziano} argued, invoking
cluster decomposition, 
that the $N^{\rm th}$ root of this expression is 
$G = \langle \lambda \lambda \rangle$.   

It is known
that the single instanton computation makes an order one error in these quantities.  The corrections can be understood as dilute gas corrections (in the sense that they can
be shown to arise from the sector with topological number one~\cite{hollowood}).  If the naive reasoning were correct, these effects would be
suppressed by further powers of $e^{-N}$, but this is not the case.  This is consistent with the infrared cutoff computations
suggested in~\cite{dineetalshifman}.

\section{Speculations on Real QCD}
\label{realqcd}

We have seen that instantons and large $N$ behavior are not necessarily incompatible, and emphasized that the appearance of branches in supersymmetric QCD is associated with the spontaneous 
breaking of a discrete symmetry.  As we take the soft breakings large, most of the $N$
vacua might disappear, leading to what we have called ``instanton" behavior.  On the other hand, given that the
lifetimes of the states scale as $e^{-N^4}$ (in the region over which we have control), they might survive.

\subsection{Spontaneous breaking of an explicitly broken discrete symmetry}
\label{spurioussymmetries}

In this brief section, we describe the possible behaviors in terms of the realization of a spurious symmetry.
At the level of the classical action, the softly broken supersymmetric theory exhibits a symmetry with $m_{\lambda}$ viewed as a spurion:
\beq
\lambda \lambda \rightarrow e^{2 \pi i k \over N} \lambda \lambda~~~~m_\lambda \rightarrow e^{-{2 \pi i k \over N} }m_\lambda.
\eeq
If $E(m_\lambda) = E(\vert m_\lambda \vert,m_\lambda^N)$, this spurious symmetry is not spontaneously broken.  If $E(m_{\lambda})$ is not invariant under
$m_\lambda \rightarrow e^{-{2 \pi i k \over N}} m_\lambda$, however, spontaneous symmetry breaking has occurred.  This is the option realized in SBQCD, and is associated with  $N$ stationary points of the vacuum energy.
$E$ has an imaginary part outside a finite range of $\alpha=\arg m_\lambda$. 

The  existence of branches in real QCD can be mapped to  the question of whether the spurious symmetry is broken
or unbroken as $m_\lambda$
becomes much larger than $\Lambda$.   As $m_\lambda \rightarrow \infty$ and $\lambda$ is integrated out, we generate
$\theta = \arg(m_\lambda) N$.  The question is:  does $E$ behave (in the pure gauge theory) as a function
of $\arg(m_\lambda) $ or
$\arg(m_\lambda) N$?
Needless to say, analytic tools to address this question
are not available, but we can look to
toy models to gain some understanding of the possibilities. 

We can illustrate these possible behaviors of the pure gauge theory in a field theory of scalars, treating the system classically and including
certain non-renormalizable couplings.  With a complex field, $\phi$, the potential
\beq
V(\phi) = -\mu^2 \vert \phi \vert^2 + {\lambda \over 2} \vert \phi\vert^4  - \Gamma (\phi^N + \phi^{*N})
\eeq
respects a $Z_N$ symmetry.  If $\Gamma$ is small, we can write:
\beq
\phi =f e^{ia/f}\,, ~~~f =  \sqrt{\mu^2 \over \lambda}
\eeq
The field $a$ acquires a potential
\beq
V(a) =- \Gamma f^N\cos\left(N{a \over f}\right).
\eeq
The system has $N$ degenerate minima, at ${a \over f} = {2 \pi k \over N}$, reflecting 
the spontaneous breaking of the discrete symmetry.

Adding a coupling
\beq
\delta V = m_\lambda \Lambda^2 \phi + \rm c.c.
\eeq
breaks the $Z_N$ symmetry explicitly, and the parameter $m_\lambda$ is a spurion
analogous to $m_\lambda$ in SUSY QCD.    For small $m_\lambda= \vert m_\lambda \vert e^{i\alpha}$, $\phi$ does not shift significantly, and
the classical vacuum energy has a contribution
\beq
E(\alpha,k) = \vert m_\lambda\vert  \Lambda^2 f \cos\left(\alpha + {2 \pi k \over N}\right).
\eeq
The potential reflects the spontaneous breaking of the spurious symmetry. Quantum mechanically, $E$ has a small imaginary part except for $k$ such that $\vert \alpha + {2 \pi k \over N}\vert < \pi.$

Elsewhere in the parameter space, however, the branches disappear.  For example, for $\mu^2$ negative, the potential
has a unique minimum, and this is not altered by the addition of the $m_\lambda$ term.  Instead,
\beq
\langle \phi \rangle = {m_\lambda^* \Lambda^2\over  \mu^2},
\eeq
and
\beq
E(\alpha,k) = { |m_\lambda|^2 \Lambda^4 \over \mu^2}.
\eeq
Thinking of this as a toy model of supersymmetric QCD, the parameters $\mu^2 \rightarrow \mu^2(m_\lambda)$,
$\Gamma \rightarrow \Gamma(m_\lambda)$.  If, for example, $\mu^2(m_\lambda)$ becomes negative
and $\Gamma$ does not grow too rapidly for large $m_\lambda$, the branched structure disappears.  Alternatively, if for large $m_\lambda$, $\mu^2 >0$ and if $\Gamma$ grows
rapidly with $m_\lambda$, then the branched structure survives.  In this toy model, the $N$ vacua reflect an approximate $Z_N$ symmetry which survives in the limit.

\subsection{Stability of Branches}

In SBQCD, both with and without matter, we can ask about the stability of different branches.  Take $k=0$ and $0<\theta<2 \pi$ and consider what happens as $\eta^\prime$ increases.  At some point, the state with $k= -1$ has lower energy, and the system can
tunnel.  For small $m_\lambda$, the tunneling rate is highly suppressed, roughly as\footnote{This estimate appears also in \cite{shifmanbranches}, which notes that due to numerical factors, even for $m_\lambda \Lambda_{QCD}$, the states may be short-lived unless $ N> 100$ or so.  If true, it would be challenging to understand how
the large $N$ limit could be valid for $N \approx 3$.}, 
\beq
\Gamma = C e^{-a N^4{\Lambda^3 \over m_\lambda^3}}.
\label{klifetime}
\eeq
We can repeat this for larger $k$, producing a large set of metastable states. 
Increasing $m_\lambda$, eventually we can no longer perform a reliable computation, but based on~(\ref{klifetime}) it is possible that tunneling rates remain exponentially suppressed with $N$.  The presence of metastable states in QCD is an interesting target for the lattice~\cite{teperreview}, and could conceivably have implications for
physics in the early universe.

\subsection{'t Hooft's Picture of Confinement:  A Candidate Setting for Branched Structure}

Nambu, Mandelstam, and 't Hooft suggested that condensation of magnetically charged objects in a non-abelian theory could account for confinement of color charge~\cite{Nambu:1974zg,Mandelstam:1974pi,'tHooft:1975pu}.  
Subsequently, 't Hooft studied the adjoint-valued composite field $\Phi=F_{\mu \nu} \tilde F^{\mu \nu}$, choosing a gauge in which $\Phi$ is  everywhere diagonal and leaving unfixed a $U(1)^{N-1}$ symmetry~\cite{oblique}. He speculated that
singular points with respect to the gauge choice
correspond to massless, condensing monopoles of the $U(1)^{N-1}$ theory, and noted that in the presence of $\theta$, the monopoles  acquire a charge through the Witten effect.  When $\theta\rightarrow\theta+2 \pi$,  the spectrum is the same, but ``rearranged": what were monopoles with one charge at $\theta=0$ become monopoles of a different charge at $\theta = 2\pi$.  This picture of confinement thus gives rise to an explicit realization of branched structure with $\theta$.  The details,
including whether there are $N$ vacua of a spurious $Z_N$ symmetry, depend on unknown features of the monopole/dyon spectrum.
Such dynamical features are also suggested by consideration of the algebra of Wilson and 't Hooft lines~\cite{thooftlines,ast}.

$N=2$ supersymmetric Yang-Mills, with a small mass $m_A$ for the adjoint
chiral multiplet,  exhibits many of these features explicitly, including a $U(1)^{N-1}$ symmetry in the small $m_A$ limit.  Seiberg and Witten showed that
the theory possesses massless monopoles at points in the moduli space~\cite{seibergwitten}. In the case of $SU(N)$, there are $N$ such points,
related by a discrete $Z_N$ symmetry.    Turning on $m_A$, the massless monopoles condense, and the theory confines.  The condensate is proportional to $m_A$, and, for small $m_A$,  the monopole and $U(1)$ gauge field masses are also suppressed by $m_A$.  The theory possesses precisely the sort of branched structure  anticipated by 't Hooft, with
$\tau = {8 \pi^2 \over g^2} + ia$, and the branches are associated with
the $Z_N$ symmetry of the theory. As $m_A$ becomes larger than $\Lambda$, it is not clear what becomes of the monopole picture; the $U(1)$ gauge bosons are no longer light relative to other states in the spectrum, nor are the
monopoles.  But we know that the $N=1$ theory exhibits a branched structure.  

If 't Hooft's picture for confinement is qualitatively correct for real QCD, it can account for a branched structure. However, the applicability of the monopole condensation picture to real QCD remains unclear. For example, one does not expect that the theory exhibits light states corresponding to $U(1)^{N-1}$ gauge bosons.  Starting from $N=2$, it is also not clear that the monopole picture is instructive for large $m_A$, let alone after adding a soft breaking gaugino mass.

\section{Summary}%:  Two Alternative Possibilities for the $\theta$-Dependence of Large N QCD}
\label{conclusions}

We have studied the large $N$ $\theta$ dependence of supersymmetric QCD, using small soft breakings as a probe of the nonsupersymmetric limit. We have seen that certain aspects of the usual large $N$ picture, including the presence of branches and the behavior in theories with matter (both with $N_f \ll N$ and $N_f \sim N$), are reflected in SBQCD. However, there are also striking departures from ordinary QCD and the conventional large $N$ description. First, in supersymmetric theories, instanton effects
are sometimes calculable and do not fall off exponentially with $N$.  Second, branched structure in SBQCD is always associated with approximate discrete symmetries, which are badly broken in the nonsupersymmetric limit.

In light of these differences, and to advance our understanding of nonperturbative phenomena in QCD, it would be of great interest to have additional lattice probes of the branched structure of large $N$ QCD.  In future work we will explore aspects of lattice tests, particularly the possibility of searching directly for the tower of metastable states at $\theta=0$.

\vskip 1cm
\noindent
{\bf Acknowledgements:}  This work was supported in part by the U.S. Department of Energy grant number DE-FG02-04ER41286. 
We are grateful for conversations with and critical comments from Tom Banks,  Nathan Seiberg, Steve Shenker, and Edward Witten.

\vskip 1cm

\section*{\bf  Appendix:  $\Lambda$ and $\Lambda_{hol}$}

Quantities in supersymmetric gauge theories are readily derived in terms of an object referred to as the {\it holomorphic
scale}, $\Lambda_{hol}$.  In the case of $SU(N)$ SUSY QCD without chiral fields, we can make this notion precise in a very
simple way, embedding the theory in an ${\cal N}=4$  theory, with masses for the adjoint fields providing a cutoff for the SQCD theory~\cite{arkanihamedmurayama,dineetalshifman}.  In a presentation in which the $SU(4)$ symmetry is (almost) manifest, the
action is
\beq
{\cal L} = -{1 \over 32 \pi^2} \int d^2 \theta \tau W_\alpha^2  + {1 \over g^2} \int d^4 \theta \Phi_i^\dagger e^V \Phi^i
+ \int d^2 \theta {1 \over g^2} f_{abc} \epsilon^{ijk} \Phi^a_i \Phi^b_j \Phi^c_k.
\eeq
Here $\tau$ is
\beq
\tau = {8 \pi^2 \over g^2} + i \theta.
\eeq
In order that the superpotential be a holomorphic function of $\tau$, we rescale the $\Phi^a$ fields.  We can also add holomorphic mass terms:
\beq
{\cal L} = -{1 \over 32 \pi^2} \int d^2 \theta \tau W_\alpha^2  + {1 \over g^{2/3} }\int d^4 \theta \Phi_i^\dagger e^V \Phi^i
+ \int d^2 \theta ( f_{abc} \epsilon^{ijk} \Phi^a_i \Phi^b_j \Phi^c_k + M \Phi^a_i \Phi^a_i).
\eeq
Holomorphy of the gauge coupling function gives, for the renormalized coupling,
\beq
{8 \pi^2 \over g^2(m)} = {8 \pi^2 \over g^2(M)} + b_0 \log(m/M) .
\eeq  
Here $m$ and $M$ are holomorphic parameters (this is discussed further in \cite{dineetalshifman}).
The physical masses are related to these by a factor of $g^{2/3}(m), g^{2/3}(M)$; substituting yields the
standard $\beta$ function through two loops (issues involving the exact $\beta$ function are discussed,
again, in \cite{dineetalshifman}).  $\Lambda_{hol}$ is then defined through:
\beq
\Lambda_{hol} = M e^{-\tau/b_0} = g^{-2/3} M_{phys} e^{-\tau/b_0} .
\eeq
 This is {\it almost} the conventionally defined $\Lambda$ parameter, but in large $N$ it differs by a power of $N$, as noted in \cite{shifmannscaling} and we now review.

The Particle Data Group presents the strong coupling as (with slight redefinition of $b_0$ and $b_1$
to agree with our conventions above):
\beq
\alpha_s(\mu) = {4 \pi \over b_0 t} \left (1 - {b_1 \over b_0^2} {\log t \over t } \right ),~~~t = \log\left({\mu^2 \over \Lambda^2}\right).
\label{pdgalphas}
\eeq
Comparing with the solution of the RGE,
\beq
{8 \pi^2 \over g^2(\mu)} = {8 \pi^2 \over g^2(M_{phys})} + b_0 \log(\mu/M_{phys}) - {b_1 \over b_0} \log(g(\mu)/g(M_{phys})),
\label{eq:twolooprge}
\eeq  
we see that inserting
\beq
\Lambda = M_{phys} e^{-{8 \pi^2 \over b_0 g^2(M_{phys})} } \left ( \sqrt{b_0 \over 8 \pi^2} g(M_{phys}) \right )^{-b_1/b_0^2},
\eeq
and 
\beq
\log t \approx \log\left({8 \pi^2 \over b_0 g^2(\mu)}\right)
\eeq
into Eq.~(\ref{pdgalphas}), we recover Eq.~(\ref{eq:twolooprge}). Using $b_1^2/b_0=2/3$ in pure SYM, one obtains $(\Lambda_{hol}/\Lambda)^3\sim N$ in large $N$.

\bibliography{monodromy_in_field_theory}{}
\bibliographystyle{utphys}

\end{document}